\documentclass[aps,prl,floatfix,twocolumn]{revtex4}
\usepackage{graphicx}
\graphicspath{{figures/}}

\usepackage{units,xspace}
\newcommand{\micron}{\ensuremath{\unit{\mu m}}\xspace}

\begin{document}

\title{Microoptomechanical pumps assembled and driven by holographic optical vortex arrays}

\author{Kosta Ladavac}

\affiliation{James Franck Institute and Dept. of Physics\\
The University of Chicago, Chicago, IL 60637}

\author{David G. Grier}
\affiliation{Dept. of Physics and Center for Soft Matter Research\\
New York University, New York, NY 10003}

\date{\today}

\begin{abstract}
  Beams of light with helical wavefronts can be focused into
  ring-like optical traps known as optical vortices.
  The orbital angular momentum carried by photons in helical modes
  can be transferred to trapped mesoscopic objects and thereby coupled
  to a surrounding fluid.
  We demonstrate that arrays of optical vortices created with the
  holographic optical tweezer technique can assemble colloidal
  spheres into dynamically reconfigurable 
  microoptomechanical pumps assembled by optical gradient
  forces and actuated by photon orbital angular momentum.
\end{abstract}

\pacs{XXX}

\maketitle

The ever-shrinking scale and increasing complexity of microfluidic systems 
has created a need for new
methods to pump and steer fluids through
micrometer-scale
channels.
Approaches based on hydraulic control \cite{unger00} and 
electroosmosis \cite{bousse00} are ideal for many applications
and can be implemented flexibly with rapid prototyping methods, and
mass-produced with lithographic techniques.
They require external control apparatus, however, and
are not easily reconfigured in real time.
Elegant microfluidic pumps created by driving
colloidal particles with actively scanned optical tweezers \cite{terray02}
require \emph{in situ} assembly and so do not lend themselves to
low-cost or highly integrated systems.

This Letter describes a new approach to
microfluidic control based on the properties of generalized 
optical traps known as optical vortices \cite{he95,simpson96,gahagan96}
 that are capable of 
exerting torques as well as forces.  In particular, we employ
the recently introduced holographic optical tweezer technique
\cite{dufresne98}
to create arrays of optical vortices that organize fluid-borne 
colloidal particles into rapidly circulating rings, thereby
generating fluid flows with pinpoint control and no moving parts.

Our method builds on the insight \cite{allen99} that
helical modes of light
carry orbital angular momentum
that can be transferred to illuminated objects 
\cite{he95a,friese96,oneil02,curtis03}.
Strongly focusing such a beam with a high numerical aperture lens creates
a variant of an optical tweezer \cite{ashkin86} known variously
as an optical vortex, optical spanner, or optical wrench \cite{he95,simpson96,gahagan96}.
The necessary helical modes are easily generated from conventional
Gaussian TEM$_{00}$ beams with computer-designed diffractive
mode converters \cite{he95}
designed to imprint the helical phase function
$\exp(i \ell \theta)$ onto the light's wavefronts.
Here, $\theta$ is the azimuthal angle around the beam's axis,
and $\ell$ is an integer winding number describing the helix's
pitch. 
Such a beam focuses to a ring of light rather than a bright spot
because destructive interference cancels the beam's
intensity along its axis.
The radius of the dark central core increases 
linearly with $\ell$ \cite{curtis03}.
\begin{figure}[t!]
  \centering
  \includegraphics[width=\columnwidth]{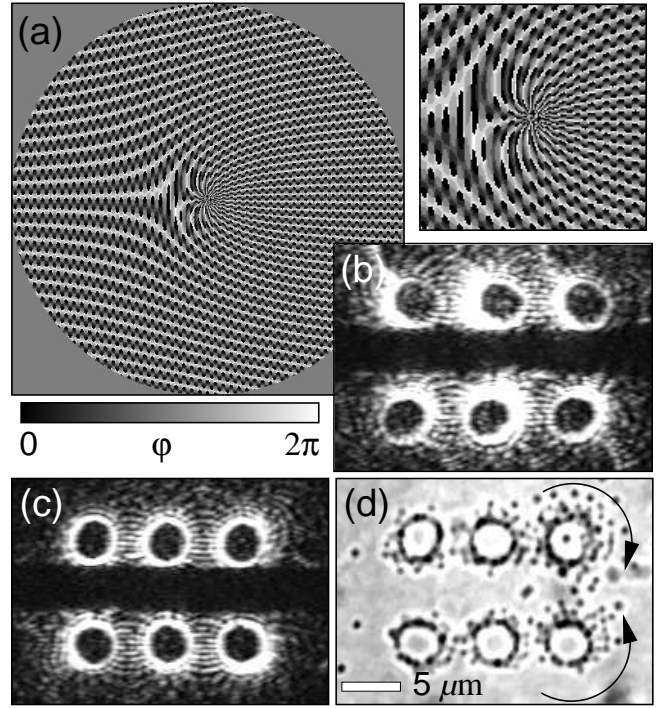}
  \caption{Creating a microfluidic pump from a beam of light.
    (a) Gray-scale representation of the
    phase hologram, $\varphi(\vec{r})$, encoding an
    optomechanical pump.  The color bar translates gray-scale to phase
    shifts in radians.  
    The inset is an expanded view near the singularity
    at the optical axis.
    (b) Focused image of the
    $3 \times 2$ optical vortex array projected by $\varphi(\vec{r})$.  
    (c) The same optical vortex array after aberration correction.
    (d) Bright-field image of 800~\unit{nm}
    diameter silica spheres trapped in the array of 
    optimized optical vortices.
  }
  \label{fig:light}
\end{figure}

Dielectric objects comparable in size to the wavelength of light
are drawn by optical gradient forces toward an optical vortex's
bright ring, and are driven around its circumference
by the tangential component of the beam's momentum flux.
Colloidal particles dispersed in a viscous fluid can be stably
trapped near the focal plane.
Their circulation around the ring entrains flows
in the surrounding fluid that can be harnessed for controlled
transport at extremely small scales.

A single optical vortex does little more than stir a micrometer-scale
volume.  Arrays of optical vortices, however, can excite
larger flows.  We create such arrays using
the holographic optical tweezer technique \cite{dufresne98,curtis02}
in which a computer-generated hologram splits a single laser
beam into multiple independent beams, each of which can be focused
into a separate optical trap.
Each diffracted beam, moreover, can be transformed by the same hologram
into a helical mode with an individually specified winding number, $\ell$ 
\cite{curtis02}.
The phase hologram, $\varphi(\vec{r})$, shown in Fig.~\ref{fig:light}(a)
encodes the $3 \times 2$ array of optical vortices whose focal waists
appear in Figs.~\ref{fig:light}(b) and \ref{fig:light}(c).  
The upper row of optical vortices has topological charge $\ell = +21$ and the lower has
opposite helicity, $\ell = - 21$.  The two rows therefore exert torques
with opposite senses.

We use a liquid crystal phase-only spatial light modulator (SLM) (Hamamatsu
X7550 PAL-SLM) to imprint the trap-forming phase pattern, $\varphi(\vec{r})$,
onto the collimated 
beam provided by a diode-pumped frequency-doubled Nd:YVO$_4$ 
laser (Coherent Verdi) 
operating at $\lambda = 532~\unit{nm}$.
The SLM can selectively shift the light's phase
between 0 and $2 \pi$ radians with 150 calibrated phase gradations
at each 40~\micron wide pixel in a $480 \times 480$ array.
Positioning the SLM in a plane conjugate to the input pupil of
a $100 \times$ NA 1.4 S-Plan Apochromat oil-immersion objective lens
ensures that each beam diffracted
by the phase modulation imposed by the SLM passes through the input
pupil and forms an optical trap.
The resulting intensity distribution shown in Fig.~\ref{fig:light}(b)
was imaged
by placing a mirror in the objective's focal plane and collecting the
reflected light with a monochrome CCD camera.
A linear spatial filter aligned with the pump's axis in an intermediate focal plane 
blocks the
undiffracted portion of the input beam, which otherwise would create
a strong optical tweezer in the middle of the field of view.

Optical vortices are very sensitive to aberrations both in their
shape and also in the distribution of light around their circumference.
Brightness variations are particularly problematic because particles
tend to become localized by optical gradient
forces in the brightest regions \cite{curtis03}.
As little as $\lambda/10$ of coma can prevent a single particle
from circulating around an optical vortex.
These distortions are exacerbated in arrays of optical traps created
with non-ideal phase masks, so that even a well-aligned optical train
results in warped, nonuniform rings such as those in
Fig.~\ref{fig:light}(b).
Fortunately, we can correct for the measured aberrations in our optical train
\cite{born99}
by appropriately modifying $\varphi(\vec{r})$.
The optimized phase profile
combines the functions of a beam-splitter,
a mode-former, and an adaptive optical wavefront corrector
and yields the uniform optical vortices in Fig.~\ref{fig:light}(c),
each of which can induce a single particle to circulate freely
\cite{curtis03}.

We projected this trap array into a colloidal dispersion of 800~\unit{nm}
diameter silica spheres 
(Bangs Laboratories catalog number SS03N)
dispersed in a 12~\micron thick layer of water between a glass microscope
slide and a \#1 cover slip.
Spheres are drawn to the focal rings and immediately begin circulating,
with those in the upper row cycling clockwise and those in the lower row
moving counterclockwise.
To prevent particles escaping from the rings along the axial direction,
we focused the optical vortex array about $h = 2~\micron$ below the
upper glass surface.
A typical snapshot of the optically organized structure
appears in Fig.~\ref{fig:light}(d).
This is the first demonstration of
motion driven by a heterogeneous array of optical vortices.

\begin{figure}[htbp]
  \centering
  \includegraphics[width=0.9\columnwidth]{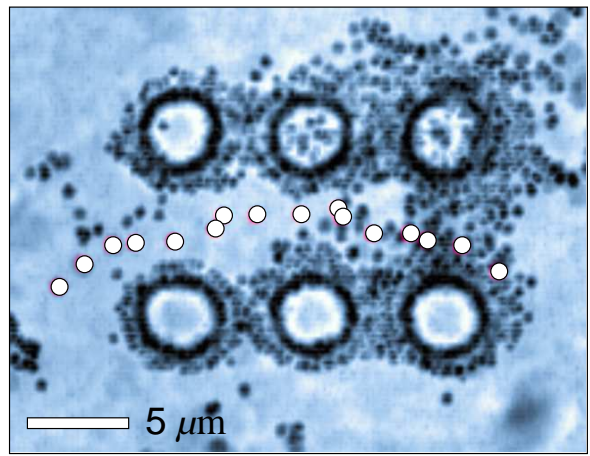}
  \caption{Time-lapse composite of 16 images in half-second intervals
    of colloidal spheres in the holographic pump at $P = 2.4~\unit{W}$.
    Circles identify the trajectory of a single sphere as it moves
    25~\micron to the left in 7~\unit{sec}.  Its peak speed is 
  $5~\unit{\micron/sec}$.}
  \label{fig:multi}
\end{figure}

Each ring in Fig.~\ref{fig:light}(d) has a radius of 
$R = 1.9 \pm 0.1~\micron$.
The rings' centers are separated by $6.6 \pm 0.2~\micron$ 
in each row, and the rows are
separated by $9.1 \pm 0.2~\micron$.
The entire pattern is centered symmetrically on the optical axis.
All of these dimensions, including the position and geometry of the array,
are easily changed by appropriately modifying $\varphi(\vec{r})$ \cite{curtis02}.
The choice of winding number,
$\ell = \pm 21$, was found to optimize the optomechanical
efficiency in our apparatus \cite{curtis03}. 

\begin{figure}[t!]
  \centering
  \includegraphics[width=\columnwidth]{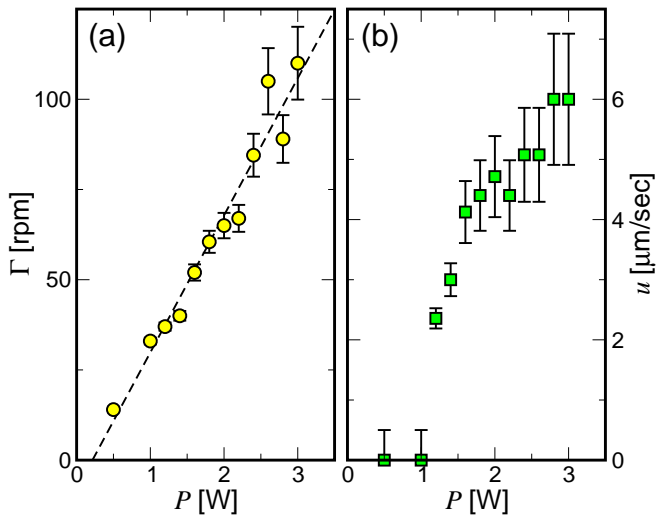}
  \caption{(a) Circulation rate in revolutions per minute (rpm) and 
    (b) axial flow speed, as a function of laser power.}
  \label{fig:trends}
\end{figure}

Coordinated counter-rotation of the two ranks of optically-trapped
spheres drives a steady flow from right to left in our images, 
through the 5~\micron wide clear channel between the rows,
with a return flow outside the pattern.
Particles that are not trapped in the optical vortices serve as passive
tracers of the fluid flow.
Figure~\ref{fig:multi} shows a multiply exposed image of colloidal spheres'
interaction with the optical vortex array.
The majority of spheres drawn into the pump's inlet on the right
become trapped in the rings.
Some wander into the dark central channel, where they are advected
by the flowing water.
The circles in Fig.~\ref{fig:multi} mark the trajectory of one such sphere.
We analyzed recorded video sequences to measure 
the trapped spheres' circulation rate, $\Gamma(P)$, 
and the flowing spheres' speed, $u(P)$, as a
function of laser power.
The results appear in Fig.~\ref{fig:trends}.

Each data point in Fig.~\ref{fig:trends}(a) is an average over all six rings, with
error bars reflecting both the roughly 5\% variation from ring to ring
and the estimated $0.05~\unit{sec}$ measurement error
in the circulation period.
The maximum circulation rate, $\Gamma = 1.8 \pm 0.2~\unit{Hz}$, attained at the highest operating
power of our SLM, $P = 3~\unit{W}$, corresponds to a circumferential speed 
of $u_0(P) = 22\pm 2~\unit{\micron/sec}$.
No-slip boundary conditions at the spheres' surfaces ensure that the
fluid at each vortex's rim also flows at this speed.

The flow field elsewhere in the system can be approximated by
the superposition of flow fields due to point forces,
known as stokeslets \cite{pozrikidis92}, 
arranged in a ring of radius $R$ at height $h$ below a plane wall.
Ensuring that the flow vanishes on the wall requires us to account for
the stokeslets' hydrodynamic images in the wall \cite{blake71}.
Accounting for the drag due to the second wall would require a substantially
more sophisticated analysis \cite{pozrikidis92,liron76}
and does not substantially affect the leading-order behavior \cite{dufresne01}.
The azimuthal component of a single stokeslet ring's wall-corrected
far-field flow at radius $r$ from the core and at height $h$ below the wall
scales as
\begin{equation}
  \label{eq:flow}
  u(r) = u(R) \, \left(\frac{R}{r}\right)^2
\end{equation}
to lowest order in $R/r$.
We can apply this result to our experimental data by identifying
the spheres' observed circumferential speed, $u_0(P)$, with $u(R)$ for spheres
substantially smaller than $R$.
This prefactor automatically includes all optical and drag-induced effects.
In the same approximation, the net flow along the channel is the superposition
of single-ring flows, with inter-ring coupling also accounted for by $u_0(P)$.
Given the arrangement of rings in our system, we anticipate a
peak axial flow speed of $u(P) \approx u_0(P)/3$.

We gauge $u(P)$ by tracking particles that wander into the
shadow of the linear
beam block along the pump's axis.
The result, plotted in Fig.~\ref{fig:trends}(b), reaches
$u = 6~\unit{\micron/sec}$ at 
$P = 3~\unit{W}$,
as predicted by Eq.~(\ref{eq:flow}).
This compares favorably with the performance
of actively actuated optical peristaltic pumps
\cite{terray02} and
has the added benefit that the pumping structure and its
action are all encoded in a single \emph{static} phase pattern
that can be implemented with a low-cost 
microfabricated diffractive optical element.

The comparatively low opto-mechanical efficiency of our implementation
can be ascribed to two effects, both of which reflect practical rather
than fundamental limitations.
The first is the relatively low efficiency of our optical train,
in which only 20 percent of the laser power $P$ is actually
projected into the sample in the desired mode.
The pixellated SLM, furthermore, cannot encode the continuous
phase modulation required to create an ideal optical vortex.
As a result, the circumference of each ring suffers from an $\ell$-fold
intensity modulation
that establishes a corrugated potential energy
landscape \cite{curtis03}.
These corrugations deepen as the laser power increases to the point that
a single sphere stops circulating altogether.
Single-vortex measurements \cite{curtis03} suggest that the pump would
stall at $P \approx 12~\unit{W}$.
Below this threshold, the corrugations
increase the effective drag on the spheres, and may
diminish the circulation rate by as much as 90 percent \cite{curtis03}.
Creating holographic microoptomechanical pumps with more refined phase masks in 
higher-efficiency
optical trains therefore should yield substantial efficiency gains.

Loading an optical vortex with multiple spheres offers several benefits
for optomechanical drive.
Each sphere captures some of the orbital angular momentum flux from the light,
and thereby contributes to the linear momentum transferred to the water.
Furthermore, collisions among densely trapped spheres helps to minimize the deleterious
effects of intensity corrugations and hot spots.
Adding too many spheres, however, can jam the pump.

More nuanced and efficient diffractive optical elements would facilitate creating
larger and more sophisticated pumping configurations.
This suggests opportunities for creating dynamically reconfigured microfluidic
systems without physical walls.
Also, the chemically synthesized colloidal particles used in our demonstration
could be replaced by micromachined gears.
Optical vortices trained on the gears' teeth would set them spinning, thereby
creating optically driven gear pumps.
All of this functionality can be applied to three-dimensional devices by 
further transforming the helical modes used in this study into self-healing
diffractionless Bessel beams \cite{arlt01,garceschavez02}.

We are grateful for helpful interactions with Brian Koss and Jennifer Curtis
This work was supported by the National Science Foundation under Grant Number
DMR-0304906.


\begin{thebibliography}{21}
\expandafter\ifx\csname natexlab\endcsname\relax\def\natexlab#1{#1}\fi
\expandafter\ifx\csname bibnamefont\endcsname\relax
  \def\bibnamefont#1{#1}\fi
\expandafter\ifx\csname bibfnamefont\endcsname\relax
  \def\bibfnamefont#1{#1}\fi
\expandafter\ifx\csname citenamefont\endcsname\relax
  \def\citenamefont#1{#1}\fi
\expandafter\ifx\csname url\endcsname\relax
  \def\url#1{\texttt{#1}}\fi
\expandafter\ifx\csname urlprefix\endcsname\relax\def\urlprefix{URL }\fi
\providecommand{\bibinfo}[2]{#2}
\providecommand{\eprint}[2][]{\url{#2}}

\bibitem[{\citenamefont{Unger et~al.}(2000)\citenamefont{Unger, Chou, Thorsen,
  Scherer, and Quake}}]{unger00}
\bibinfo{author}{\bibfnamefont{M.~A.} \bibnamefont{Unger}},
  \bibinfo{author}{\bibfnamefont{H.~P.} \bibnamefont{Chou}},
  \bibinfo{author}{\bibfnamefont{T.}~\bibnamefont{Thorsen}},
  \bibinfo{author}{\bibfnamefont{A.}~\bibnamefont{Scherer}}, \bibnamefont{and}
  \bibinfo{author}{\bibfnamefont{S.~R.} \bibnamefont{Quake}},
  \bibinfo{journal}{Science} \textbf{\bibinfo{volume}{288}},
  \bibinfo{pages}{113} (\bibinfo{year}{2000}).

\bibitem[{\citenamefont{Bousse et~al.}(2000)\citenamefont{Bousse, Cohen,
  Nikiforov, Chow, Kopf-Sill, Dubrow, and Parce}}]{bousse00}
\bibinfo{author}{\bibfnamefont{L.}~\bibnamefont{Bousse}},
  \bibinfo{author}{\bibfnamefont{C.}~\bibnamefont{Cohen}},
  \bibinfo{author}{\bibfnamefont{T.}~\bibnamefont{Nikiforov}},
  \bibinfo{author}{\bibfnamefont{A.}~\bibnamefont{Chow}},
  \bibinfo{author}{\bibfnamefont{A.~R.} \bibnamefont{Kopf-Sill}},
  \bibinfo{author}{\bibfnamefont{R.}~\bibnamefont{Dubrow}}, \bibnamefont{and}
  \bibinfo{author}{\bibfnamefont{J.~W.} \bibnamefont{Parce}},
  \bibinfo{journal}{Annu. Rev. Biophysics Biomolecular Structure}
  \textbf{\bibinfo{volume}{29}}, \bibinfo{pages}{155} (\bibinfo{year}{2000}).

\bibitem[{\citenamefont{Terray et~al.}(2002)\citenamefont{Terray, Oakey, and
  Marr}}]{terray02}
\bibinfo{author}{\bibfnamefont{A.}~\bibnamefont{Terray}},
  \bibinfo{author}{\bibfnamefont{J.}~\bibnamefont{Oakey}}, \bibnamefont{and}
  \bibinfo{author}{\bibfnamefont{D.~W.~M.} \bibnamefont{Marr}},
  \bibinfo{journal}{Science} \textbf{\bibinfo{volume}{296}},
  \bibinfo{pages}{1841} (\bibinfo{year}{2002}).

\bibitem[{\citenamefont{He et~al.}(1995{\natexlab{a}})\citenamefont{He,
  Heckenberg, and Rubinsztein-Dunlop}}]{he95}
\bibinfo{author}{\bibfnamefont{H.}~\bibnamefont{He}},
  \bibinfo{author}{\bibfnamefont{N.~R.} \bibnamefont{Heckenberg}},
  \bibnamefont{and}
  \bibinfo{author}{\bibfnamefont{H.}~\bibnamefont{Rubinsztein-Dunlop}},
  \bibinfo{journal}{J. Mod. Opt.} \textbf{\bibinfo{volume}{42}},
  \bibinfo{pages}{217} (\bibinfo{year}{1995}{\natexlab{a}}).

\bibitem[{\citenamefont{Gahagan and Swartzlander}(1996)}]{gahagan96}
\bibinfo{author}{\bibfnamefont{K.~T.} \bibnamefont{Gahagan}} \bibnamefont{and}
  \bibinfo{author}{\bibfnamefont{G.~A.} \bibnamefont{Swartzlander}},
  \bibinfo{journal}{Opt. Lett.} \textbf{\bibinfo{volume}{21}},
  \bibinfo{pages}{827} (\bibinfo{year}{1996}).

\bibitem[{\citenamefont{Simpson et~al.}(1996)\citenamefont{Simpson, Allen, and
  Padgett}}]{simpson96}
\bibinfo{author}{\bibfnamefont{N.~B.} \bibnamefont{Simpson}},
  \bibinfo{author}{\bibfnamefont{L.}~\bibnamefont{Allen}}, \bibnamefont{and}
  \bibinfo{author}{\bibfnamefont{M.~J.} \bibnamefont{Padgett}},
  \bibinfo{journal}{J. Mod. Opt.} \textbf{\bibinfo{volume}{43}},
  \bibinfo{pages}{2485} (\bibinfo{year}{1996}).

\bibitem[{\citenamefont{Dufresne and Grier}(1998)}]{dufresne98}
\bibinfo{author}{\bibfnamefont{E.~R.} \bibnamefont{Dufresne}} \bibnamefont{and}
  \bibinfo{author}{\bibfnamefont{D.~G.} \bibnamefont{Grier}},
  \bibinfo{journal}{Rev. Sci. Instr.} \textbf{\bibinfo{volume}{69}},
  \bibinfo{pages}{1974} (\bibinfo{year}{1998}).

\bibitem[{\citenamefont{Allen et~al.}(1999)\citenamefont{Allen, Padgett, and
  Babiker}}]{allen99}
\bibinfo{author}{\bibfnamefont{L.}~\bibnamefont{Allen}},
  \bibinfo{author}{\bibfnamefont{M.~J.} \bibnamefont{Padgett}},
  \bibnamefont{and} \bibinfo{author}{\bibfnamefont{M.}~\bibnamefont{Babiker}},
  \bibinfo{journal}{Progr. Opt.} \textbf{\bibinfo{volume}{39}},
  \bibinfo{pages}{291} (\bibinfo{year}{1999}).

\bibitem[{\citenamefont{He et~al.}(1995{\natexlab{b}})\citenamefont{He, Friese,
  Heckenberg, and Rubinsztein-Dunlop}}]{he95a}
\bibinfo{author}{\bibfnamefont{H.}~\bibnamefont{He}},
  \bibinfo{author}{\bibfnamefont{M.~E.~J.} \bibnamefont{Friese}},
  \bibinfo{author}{\bibfnamefont{N.~R.} \bibnamefont{Heckenberg}},
  \bibnamefont{and}
  \bibinfo{author}{\bibfnamefont{H.}~\bibnamefont{Rubinsztein-Dunlop}},
  \bibinfo{journal}{Phys. Rev. Lett.} \textbf{\bibinfo{volume}{75}},
  \bibinfo{pages}{826} (\bibinfo{year}{1995}{\natexlab{b}}).

\bibitem[{\citenamefont{O'Neil et~al.}(2002)\citenamefont{O'Neil, MacVicar,
  Allen, and Padgett}}]{oneil02}
\bibinfo{author}{\bibfnamefont{A.~T.} \bibnamefont{O'Neil}},
  \bibinfo{author}{\bibfnamefont{I.}~\bibnamefont{MacVicar}},
  \bibinfo{author}{\bibfnamefont{L.}~\bibnamefont{Allen}}, \bibnamefont{and}
  \bibinfo{author}{\bibfnamefont{M.~J.} \bibnamefont{Padgett}},
  \bibinfo{journal}{Phys. Rev. Lett.} \textbf{\bibinfo{volume}{88}},
  \bibinfo{pages}{053601} (\bibinfo{year}{2002}).

\bibitem[{\citenamefont{Curtis and Grier}(2003)}]{curtis03}
\bibinfo{author}{\bibfnamefont{J.~E.} \bibnamefont{Curtis}} \bibnamefont{and}
  \bibinfo{author}{\bibfnamefont{D.~G.} \bibnamefont{Grier}},
  \bibinfo{journal}{Phys. Rev. Lett.} \textbf{\bibinfo{volume}{90}},
  \bibinfo{pages}{133901} (\bibinfo{year}{2003}).

\bibitem[{\citenamefont{Friese et~al.}(1996)\citenamefont{Friese, Enger,
  Rubinsztein-Dunlop, and Heckenberg}}]{friese96}
\bibinfo{author}{\bibfnamefont{M.~E.~J.} \bibnamefont{Friese}},
  \bibinfo{author}{\bibfnamefont{J.}~\bibnamefont{Enger}},
  \bibinfo{author}{\bibfnamefont{H.}~\bibnamefont{Rubinsztein-Dunlop}},
  \bibnamefont{and} \bibinfo{author}{\bibfnamefont{N.~R.}
  \bibnamefont{Heckenberg}}, \bibinfo{journal}{Phys. Rev. A}
  \textbf{\bibinfo{volume}{54}}, \bibinfo{pages}{1593} (\bibinfo{year}{1996}).

\bibitem[{\citenamefont{Ashkin et~al.}(1986)\citenamefont{Ashkin, Dziedzic,
  Bjorkholm, and Chu}}]{ashkin86}
\bibinfo{author}{\bibfnamefont{A.}~\bibnamefont{Ashkin}},
  \bibinfo{author}{\bibfnamefont{J.~M.} \bibnamefont{Dziedzic}},
  \bibinfo{author}{\bibfnamefont{J.~E.} \bibnamefont{Bjorkholm}},
  \bibnamefont{and} \bibinfo{author}{\bibfnamefont{S.}~\bibnamefont{Chu}},
  \bibinfo{journal}{Opt. Lett.} \textbf{\bibinfo{volume}{11}},
  \bibinfo{pages}{288} (\bibinfo{year}{1986}).

\bibitem[{\citenamefont{Curtis et~al.}(2002)\citenamefont{Curtis, Koss, and
  Grier}}]{curtis02}
\bibinfo{author}{\bibfnamefont{J.~E.} \bibnamefont{Curtis}},
  \bibinfo{author}{\bibfnamefont{B.~A.} \bibnamefont{Koss}}, \bibnamefont{and}
  \bibinfo{author}{\bibfnamefont{D.~G.} \bibnamefont{Grier}},
  \bibinfo{journal}{Opt. Comm.} \textbf{\bibinfo{volume}{207}},
  \bibinfo{pages}{169} (\bibinfo{year}{2002}).

\bibitem[{\citenamefont{Born and Wolf}(1999)}]{born99}
\bibinfo{author}{\bibfnamefont{M.}~\bibnamefont{Born}} \bibnamefont{and}
  \bibinfo{author}{\bibfnamefont{E.}~\bibnamefont{Wolf}},
  \emph{\bibinfo{title}{Principles of Optics}} (\bibinfo{publisher}{Cambridge
  University Press}, \bibinfo{address}{Cambridge}, \bibinfo{year}{1999}),
  \bibinfo{edition}{7th} ed.

\bibitem[{\citenamefont{Pozrikidis}(1992)}]{pozrikidis92}
\bibinfo{author}{\bibfnamefont{C.}~\bibnamefont{Pozrikidis}},
  \emph{\bibinfo{title}{Boundary Integral and Singularity Methods for
  Linearized Viscous Flow}} (\bibinfo{publisher}{Cambridge University Press},
  \bibinfo{address}{New York}, \bibinfo{year}{1992}).

\bibitem[{\citenamefont{Blake}(1971)}]{blake71}
\bibinfo{author}{\bibfnamefont{J.~R.} \bibnamefont{Blake}},
  \bibinfo{journal}{Proc. Cambridge Phil. Soc.} \textbf{\bibinfo{volume}{70}},
  \bibinfo{pages}{303} (\bibinfo{year}{1971}).

\bibitem[{\citenamefont{Liron and Mochon}(1976)}]{liron76}
\bibinfo{author}{\bibfnamefont{N.}~\bibnamefont{Liron}} \bibnamefont{and}
  \bibinfo{author}{\bibfnamefont{S.}~\bibnamefont{Mochon}},
  \bibinfo{journal}{J. Eng. Math.} \textbf{\bibinfo{volume}{10}},
  \bibinfo{pages}{287} (\bibinfo{year}{1976}).

\bibitem[{\citenamefont{Dufresne et~al.}(2001)\citenamefont{Dufresne, Altman,
  and Grier}}]{dufresne01}
\bibinfo{author}{\bibfnamefont{E.~R.} \bibnamefont{Dufresne}},
  \bibinfo{author}{\bibfnamefont{D.}~\bibnamefont{Altman}}, \bibnamefont{and}
  \bibinfo{author}{\bibfnamefont{D.~G.} \bibnamefont{Grier}},
  \bibinfo{journal}{Europhys. Lett.} \textbf{\bibinfo{volume}{53}},
  \bibinfo{pages}{264} (\bibinfo{year}{2001}).

\bibitem[{\citenamefont{Arlt et~al.}(2001)\citenamefont{Arlt, Garces-Chavez,
  Sibbett, and Dholakia}}]{arlt01}
\bibinfo{author}{\bibfnamefont{J.}~\bibnamefont{Arlt}},
  \bibinfo{author}{\bibfnamefont{V.}~\bibnamefont{Garces-Chavez}},
  \bibinfo{author}{\bibfnamefont{W.}~\bibnamefont{Sibbett}}, \bibnamefont{and}
  \bibinfo{author}{\bibfnamefont{K.}~\bibnamefont{Dholakia}},
  \bibinfo{journal}{Opt. Comm.} \textbf{\bibinfo{volume}{197}},
  \bibinfo{pages}{239} (\bibinfo{year}{2001}).

\bibitem[{\citenamefont{Garces-Chavez et~al.}(2002)\citenamefont{Garces-Chavez,
  McGloin, Melville, Sibbett, and Dholakia}}]{garceschavez02}
\bibinfo{author}{\bibfnamefont{V.}~\bibnamefont{Garces-Chavez}},
  \bibinfo{author}{\bibfnamefont{D.}~\bibnamefont{McGloin}},
  \bibinfo{author}{\bibfnamefont{H.}~\bibnamefont{Melville}},
  \bibinfo{author}{\bibfnamefont{W.}~\bibnamefont{Sibbett}}, \bibnamefont{and}
  \bibinfo{author}{\bibfnamefont{K.}~\bibnamefont{Dholakia}},
  \bibinfo{journal}{Nature} \textbf{\bibinfo{volume}{419}},
  \bibinfo{pages}{145} (\bibinfo{year}{2002}).

\end{thebibliography}

\end{document}